# Multi-Layer Local Graph Words for Object Recognition

Svebor Karaman[1], Jenny Benois-Pineau[1], Rémi Mégret[2] and Aurélie Bugeau[1]

[1] LaBRI - University of Bordeaux, 351, Cours de la Libération,
33405 Talence Cedex, France
`{Svebor.Karaman, Jenny.Benois-Pineau, Aurelie.Bugeau}@labri.fr`

[2] IMS - University of Bordeaux, 351, Cours de la Libération
33405 Talence Cedex, France
`Remi.Megret@ims-bordeaux.fr`

**Abstract.** In this paper, we propose a new multi-layer structural approach for the task of object based image retrieval. In our work we tackle the problem of structural organization of local features. The structural features we propose are nested multi-layered local graphs built upon sets of SURF feature points with Delaunay triangulation. A Bag-of-Visual-Words (BoVW) framework is applied on these graphs, giving birth to a Bag-of-Graph-Words representation. The multi-layer nature of the descriptors consists in scaling from trivial Delaunay graphs - isolated feature points - by increasing the number of nodes layer by layer up to graphs with maximal number of nodes. For each layer of graphs its own visual dictionary is built. The experiments conducted on the SIVAL and Caltech-101 data sets reveal that the graph features at different layers exhibit complementary performances on the same content and perform better than baseline BoVW approach. The combination of all existing layers, yields significant improvement of the object recognition performance compared to single level approaches.

## 1 Introduction

Visual object retrieval in images and videos is one of the most active fields of research. One of the most popular techniques addressing this task relies on the use of local features, e.g.using for instance keypoint based SIFT (Scale Invariant Feature Transform) of Lowe [1] or SURF (Speed-Up Robust Features) of Bay [3]. SIFT and SURF key points based descriptors are robust and discriminative local features. SIFT points are detected on local minima/maxima of a Difference of Gaussians (DoG) image computed at different scales. The SIFT key point feature is an orientation histogram in a close spatial neighborhood of the key point. SURF is based on sums of approximated Haar wavelet responses and use integral images in order to speed-up key point extraction.

In the trending approach of Bag-of-Visual-Words [2], the features are quantized in visual dictionaries by clustering and images are modeled by a distribution of the visual words within them. The Bag-of-Visual-Words approach is an adaptation of the text retrieval approach Bag-of-Words (BoW) to images. The BoVW operates on local visual features such as key points when the BoW operates on words. The semantic power of a word is much higher than which of a local key point, a visual word is also much more ambiguous than a text word. Moreover, the BoVW approach discards all spatial information about the relations between key points. Having a similar local distribution of key points in two images indicates a stronger similarity of content than sparse isolated key points.

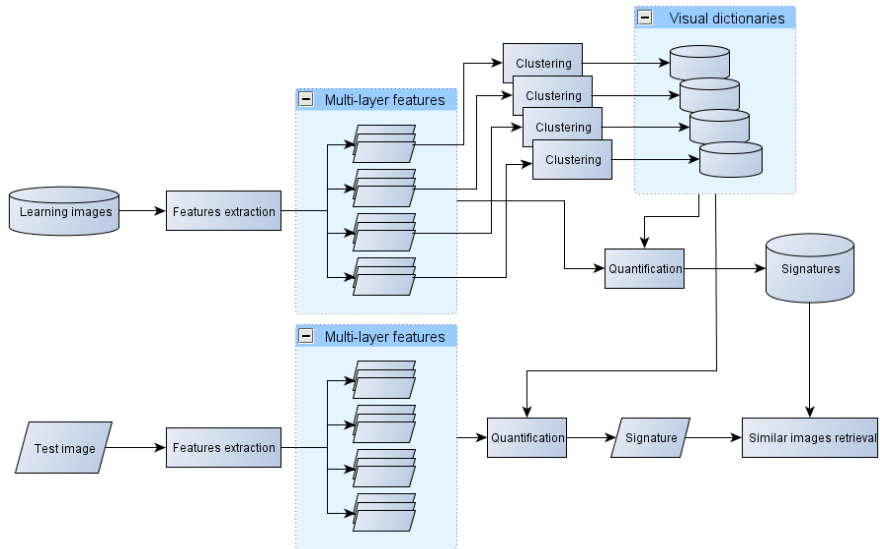

**Fig. 1.** Flowchart of the Bag-of-Words framework applied to our multi-layer features.

To overcome this limitation of the BoVW, some approaches have been developed in the past few years. The spatial pyramid matching proposed in [4] and used in [5] and [6] for its application to object or scene recognition compares the distributions on several areas generated by splitting the image spatially instead of discarding completely the geometric information by considering the image as one single distribution. However, such an approach is not invariant to affine transformations loosing the most important characteristic of points invariant local features. In [7] an approach called "Visual Phrases" is introduced to group visual words according to their proximity in the image plane as a sequence of features. The visual phrases are represented by a histogram containing the distribution of the visual words in the phrase. In these works, the common idea is to build local signature according to a visual dictionary from an arbitrary splitting for the spatial pyramid matching or on a set built by a proximity criterion for visual phrases. Compared to these works, our approach consists in introducing the local topological information within the visual features.

In this paper we propose a spatial embedding of features with local Delaunay graphs. The motivation for building such graphs comes from the invariance of Delaunay triangulation with regard to affine transformations of image plane: rotation, translation and scale. Hence, with invariant key point features such as SURF the global invariance of graphs is theoretically maintained. As such invariance is not perfect on real data, we propose to combine the structural information injected by the Delaunay graph with the robustness brought by the BoVW approach. We therefore consider multiple local Delaunay graphs as visual words, and plunge them into a Bag-of-Visual-Words framework, by building visual dictionaries obtained by clustering the sets of local graphs. Then state-of-the-art visual signatures are used for object retrieval. Increasing the number of nodes of the local graphs yields a layered approach where each layer induces a stronger spatial embedding within graph features. We call this approach "nested", as each local graph is obtained by adding nodes to a local graph from the previous layer. It combines visual signatures of all graphs from trivial graphs which are isolated SURF points to larger graphs that contain about ten nodes. The proposed framework is summarized in the flowchart presented in Figure **1**.

The paper is organized as follows, in section 2 we discuss the process of building these graphs and introduce their nested construction. In section 3, we introduce the dissimilarity measure used to compare graphs and built visual dictionaries by clustering. The latter are presented in section 4. Experiments with these new features are presented in section 5. Conclusions and perspectives are given in section 6.

## 2    Graph feature construction

Let us consider a graph ***G=(X,E)*** with ***X*** a set of nodes corresponding to some feature points $x_{k,k=1,..,K}$, in image plane and ***E**={$e_{kl}$}$_{k=1,..,K,l=1,..,K}$*, where ***$e_{kl}$**=($x_k$,$x_l$)*, a set of edges connecting these points. We call such a graph a "graph feature". We will build these features upon sets of neighboring feature points in image plane. Hence we propose a spatial embedding of local features with graphs. To build such graphs two questions have to be addressed: i) the choice of feature points sets ***X*** and ii) the design of connectivity as edges ***E***.

To define the feature point sets ***X*** upon which graphs will be built we are looking for a set of feature points that we call the "seeds". Around them, other feature points will be selected to build each graph feature. Selected seeds have to form a set of SURF points which are more likely to be detected in various instances of the same object. SURF points are detected where local maxima of the response of the approximated Hessian determinant are reached [3]. The amplitude of this criterion is a good choice for selecting the seeds, as SURF points with higher response correspond to more salient visual structures and are therefore more likely to be more repeatable. Hence, the seeds considered for building the graphs will be the SURF points with highest responses. Considering a fixed number of seeds $N_{Seeds}$, we can define the set of seeds ***S***:

$$S = \{s_1, \ldots, s_{N_{seeds}}\} \qquad (1)$$

Given ***S***, our aim is to add partial structural information of the object while keeping the discriminative power of SURF key points. We will therefore define graphs over the seeds and their neighboring SURF points.

Finding the $k$ spatial nearest SURF neighbors of each seed $s_i$ gives the set of neighbors $P_i$:

$$P_i = \{p_1, \ldots, p_k\} \qquad (2)$$

Hence the set of nodes for each graph upon a seed point is built. For the edges we use the Delaunay triangulation which is invariant with regard to affine transformations of image plane preserving angles: translation, rotation and scaling. Furthermore, regarding the future extensions of this work to video, the choice of Delaunay triangulation is also profitable for its good properties in tracking of structures [8]. The set of all vertices used for building the graph $G_i$ is $X^{Gi}$, the union of the seed and its neighborhood:

$$X^{Gi} = \{x_1^{Gi}, \ldots, x_k^{Gi}\} = P_i \cup \{s_i\} \qquad (3)$$

A Delaunay triangulation is computed on the points of $X^{Gi}$, building triangles according to the Delaunay constraint. An edge $e_{ij}=(x_i^{Gi}, x_j^{Gi})$ is defined between two vertices of the graph $G_i$ if an edge of a triangle connects these two vertices.

Introducing a layered approach, where each layer adds more structural information we can define graphs of increasing size while moving from one layer to the upper one. Each layer has his own set of neighbors around each seed $s_i$ and the Delaunay triangulation is run separately on each layer. One layer will always contain the points of all the lower layers, hence we call this approach "nested" and illustrate it in Figure **3**. To avoid a large number of layers, the number of nodes added at each layer should induce a significant change of structural information. To build a Delaunay triangulation, at least two points have to be added to the seed at the second layer. Adding one more node may yield three triangles instead of just one, resulting in a more complete local pattern. Therefore, the number of nodes added from one layer to the upper one is fixed to three. We define four layers, the bottom one containing only one SURF point, the seed, and the top one containing a graph built upon the seed and its 9 nearest neighbors, see examples in Figure **2**.

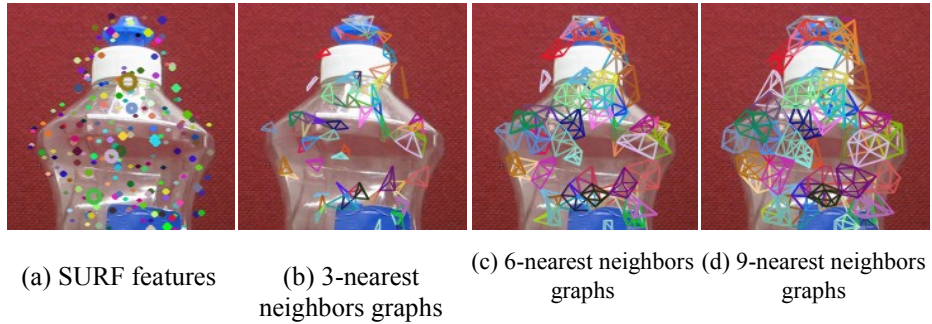

(a) SURF features    (b) 3-nearest neighbors graphs    (c) 6-nearest neighbors graphs    (d) 9-nearest neighbors graphs

**Fig. 2.** SURF and graph features on a cropped image of the object ajaxorange from SIVAL database.

## 3  Graph comparison

In order to integrate these new graph features in a Bag-of-Visual-Words framework a dissimilarity measure and a clustering method have to be defined. In this section, we define the dissimilarity measure.

We are dealing with attributed graphs, where nodes can be compared with respect to their visual appearance. Although it could be possible to take into account only similarities of node features or graph topology more information can be obtained by combining both information for defining a dissimilarity measure between local graphs. To achieve this we will investigate the use of the Context Dependent Kernel (CDK) presented in [9]. The definition of the CDK relies on two matrices: $D$ which contains the distances between node features, and $T$ which contains the topology of the graphs being compared.

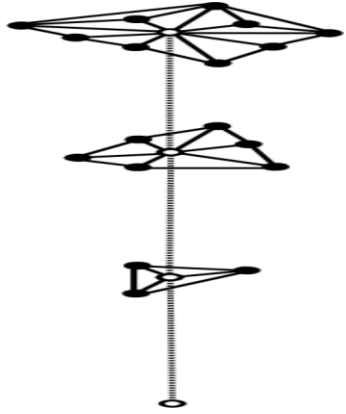

**Fig. 3.** The nested approach. Bottom to top: SURF seed depicted as the white node, 3 neighbors graph where neighbours are in black, 6 neighbors graph and 9 neighbors graph at the top level.

Considering two graphs $A$ and $B$ with respective number of nodes $m$ and $n$, let us denote $C$ the union of the two graphs:

$$C = A \oplus B$$
$$\text{with } \begin{cases} x_i^C = x_i^A & \text{for} \quad i \in [1..m] = I_A \\ x_i^C = x_{i-m}^B & \text{for} \quad i \in [m+1..m+n] = I_B \end{cases} \quad (4)$$

The feature correspondence square matrix $D$ of size $(m+n) \times (m+n)$ contains the "entrywise" $L_2$-norm (i.e., the sum of the squared values of vector coefficients) of the difference between SURF features:

$$D = (d_{ij})_{ij}$$
$$\text{where } d_{ij} = \left\| x_i^C - x_j^C \right\|_2 \quad (5)$$

The square topology matrix $T$ of size $(m+n) \times (m+n)$ defines the connectivity between two vertices $x_i^C$ and $x_j^C$. In this work we define a crisp connectivity as we set $T_{ij}$ to one if an edge connects the vertices $x_i^C$ and $x_j^C$ and 0 otherwise. Hence, only sub matrices where both lines and columns in $I_A$ or $I_B$ are not entirely null. More precisely, we can define sub matrices $T_{AA}$ and $T_{BB}$ corresponding to the topology of each graph $A$ and $B$ respectively, while sub matrices $T_{AB}$ and $T_{BA}$ are entirely null, vertices of graphs $A$ and $B$ are not connected.

$$T = (T_{ij})_{ij}$$
$$\text{where } T_{ij} = \begin{cases} 1 \text{ if edge } (x_i^C, x_j^C) \text{ belongs to } A \text{ or } B \\ 0 \text{ otherwise} \end{cases} \quad (6)$$

The CDK denoted $K$ is computed by an iterative process consisting of the propagation of the similarity in the description space according to the topology matrix.

$$K^{(0)} = \frac{\exp(-\frac{D}{\beta})}{\left\|\exp(-\frac{D}{\beta})\right\|_1} \quad , \quad K^{(t)} = \frac{G(K^{(t-1)})}{\left\|G(K^{(t-1)})\right\|_1}$$
$$G(K) = \exp\left(-\frac{D}{\beta} + \frac{\alpha}{\beta} T K^{(t-1)} T\right) \quad (7)$$

Where *exp* represents the coefficient-wise exponential and $\|M\|_1 = \Sigma_{ij}|M_{ij}|$ represents the L1 matrix norm. The two parameters **β** and **α** can be seen respectively as weights for features distance and topology propagation. Similarly to the definition of sub matrices in topology matrix $T$ we can define sub matrices in the kernel matrix $K$. The sub matrix $K_{AB}^{(t)}$ represents the strength of the inter-graph links between graphs $A$ and $B$ once the topology has been taken into account. We can therefore define the dissimilarity measure that will be used for clustering:

$$s(A, B) = \sum_{\{i \in I_A, j \in I_B\}} K_{ij}^{(t)} \in [0,1]$$
$$\rho(A, B) = s(A, A) + s(B, B) - 2s(A, B) \in [0,1] \quad (8)$$

This dissimilarity measure will be applied separately on each layer. However, for the bottom layer, since there is no topology to take into account for isolated points we will use directly the "entrywise" $L_2$-norm of the difference between SURF features.

## 4  Visual dictionaries

The state-of-the-art approach for computing the visual dictionary of a set of features is the use of the K-means clustering algorithm [2] with a large number of clusters, often several thousands. The code-word is either the center of a cluster or a non-parametric representation like a K-Nearest Neighbors (K-NN) voting approach.

Both of these approaches are not suitable for the graph-features as using the K-means clustering algorithm implies iteratively moving the cluster centers with interpolation whereas defining a mean graph is a difficult task; and a fast K-NN requires an indexing structure which is not available in our graph feature space since it is not a vector space. Therefore, we present in the following section the selected method which is a two pass agglomerative hierarchical clustering. The model of a cluster is chosen to be the median instead of the mean.

### 4.1 Clustering method

In order to quantize a very large database, it can be interesting to use a two pass clustering approach as proposed in [10], as it enables a gain in terms of computational cost. Here, the first pass of the agglomerative hierarchical clustering will be run on all the features extracted from training images of one object. The second pass is applied on clusters generated by the first pass on all objects of the database. To represent a cluster, we use the following definition of the median:

$$median = \mathrm{argmin}_{G \in V} \sum_{i=1}^{m} \|v_i - G\| \quad (9)$$

With $V$ – a cluster and $v_i$ – members of a cluster, $G$ the candidate median and $\|\cdot\|$ is a distance or dissimilarity measure in our case.

For the first pass, the dissimilarities between all the features, of the same layer, extracted on all the images of an object are computed. For the second pass, only the dissimilarities between all the medians of all object clusters are computed. Each layer being processed independently, we obtain a visual dictionary for each layer of graphs with 1, 3, .., $N_{max}$ nodes.

### 4.2 Visual signatures

The usual representation of an image in a BoVW approach is to compute a histogram of all the visual words of the dictionary within the image. Each feature extracted from an image is assigned to the closest visual word of the dictionary. We use this representation without rejection, a feature is always assigned to a word in the dictionary. The signatures are then normalized to sum to one by dividing each value by the number of features extracted from the image. Once the visual signatures of images have been computed, one can define the distance between two images as the distance between their visual signatures. In preliminary experiments we have compared results using Hamming distance, Euclidean distance and $L_1$ distance for this task. The $L_1$ distance giving better results, final results are presented using this measure only.

## 5   Experiments

The experiments are conducted on two publicly available datasets. The first one, the SIVAL (Spatially Independent, Variable Area, and Lighting) data set [11] includes 25 objects, each of them being present in 60 images taken in 10 various environment and different poses yielding a total of 1500 images. This data set is quite challenging as the objects are depicted in various lighting conditions and poses. It has also been chosen as the longer term perspective of this work is the recognition of objects of the daily living that may appear in different places of a house, for example a hoover that may be moved in all the rooms in one's house. The second one is the well known Caltech-101 [12] dataset, composed of 101 object categories. The categories are different types of animals, plants or objects. See a snippet of both datasets in Figure **4a** and **b**.

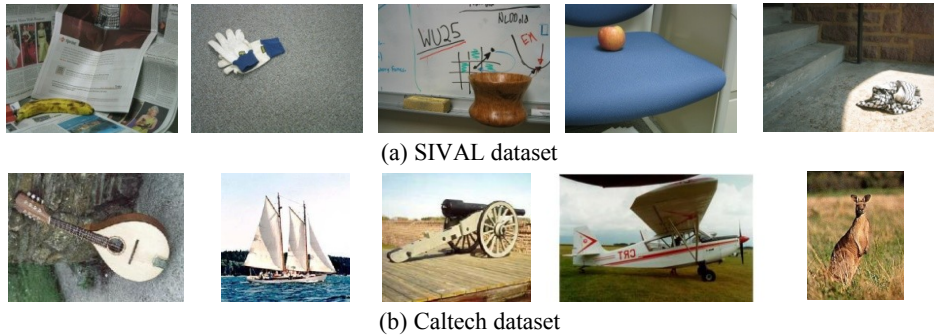

(a) SIVAL dataset

(b) Caltech dataset

**Fig. 4.** Excerpts from image datasets.

We separate learning and testing images by a random selection. On each dataset, 30 images of each category are selected as learning images for building the visual dictionaries and for the retrieval task. Some categories of Caltech-101 have several hundred of images when others have only a few more than 30. The testing images are therefore a random selection of the remaining images up to 50. We only take into account the content of a bounding box of each object as the focus of this paper is only object recognition and not yet localization. SURF key points of 64 dimensions are extracted within the bounding box, the numbers of seeds for the graphs building process is fixed to 300. The second layer corresponds to graphs built upon the seeds and their 3 nearest neighbors, the third layer with the 6 nearest neighbors and the fourth and last layer with the 9 nearest neighbors. For the CDK, α is set to 0.0001, β to 0.1 (ensuring **K** is a proper kernel) and the number of iterations is fixed to 2, as H. Sahbi [9] has shown that the convergence of the CDK is fast. The first pass clustering compute 500 clusters for each object. The final dictionary size varies in the range 50-5000. Each layer will yield its own dictionary. We compare our method with standard BoVW approach. For that purpose, we use all the SURF features available on all images of the learning database to build the BoVW dictionary. The visual words are obtained by performing k-means clustering on the set of all these descriptors. Each visual word is characterized by the center of a cluster.

The graph features are not built using all available SURF points, therefore to analyze the influence of this selection, signatures are computed for the set of SURF which have been selected to build the different layers of graphs. These configurations will be referred to as SURF3NN, SURF6NN and SURF9NN corresponding respectively to all the points upon which graphs with 3, 6 and 9 nearest neighbors have been defined. In this case the dictionaries are built with our two pass clustering approach as for graphs.

For each query image and each database image, the signatures are computed for isolated SURF and the different layers of graphs. We have investigated the combination of isolated SURF and the different layers of graphs by an early fusion of signatures i.e. concatenating the signatures. For SIVAL that concatenation has been done with the signature from the selected SURF corresponding to the highest level whereas for Caltech-101 we use the classical BoW SURF signature. Finally, the L1-distance between histograms is computed to compare two images. The performance is evaluated by the Mean Average Precision (MAP) measure. For each test images, all images in the learning set are ranked from the closest (in terms of $L_1$ distance between visual signatures) to the furthest. The average precision is evaluated for each test image of an object, and the MAP is the mean of these values for all the images of an object in the test set. For the whole database we measure the performance by the average value of the MAP i.e. we do not weight the MAP per class by the number of query as this would induce more consideration to categories with more testing.

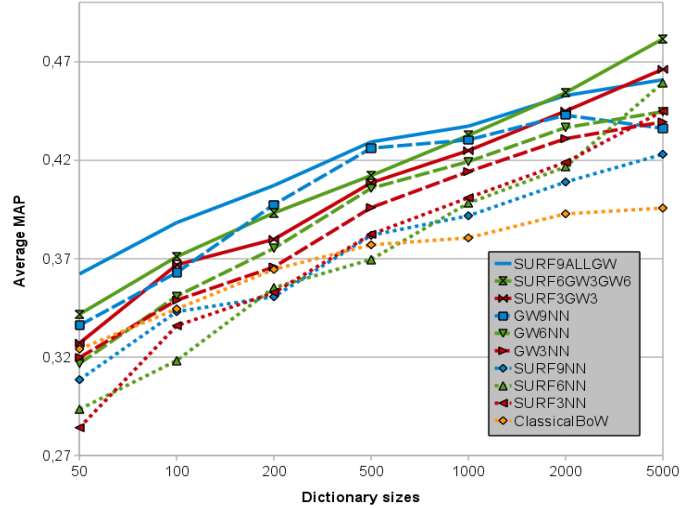

**Fig. 5.** Average MAP on the whole SIVAL data set. Isolated SURF features are the dotted curves, single layer Graphs Words are drawn as dashed curves and the multi-layer approach in solid curves.

### 5.1 SURF based BoW vs Graphs Words

First of all, it is interesting to analyze if the graph words approach where each layer is taken into consideration separately obtains similar performances compared to the classical BoVW approach using only SURF features. This is depicted in Figure **6** and **7** where isolated SURF points are depicted as dotted lines, single layer of graph words are dashed lines and the combination of SURF and different graphs layers are plotted as continuous lines. At first glance, we can see that for SIVAL isolated SURF features perform the poorest, separated layers of graphs performs better and the combination of different layers of graphs and the SURF features upon which the highest layer have been computed obtain the best performances. Our clustering approach seems to give worst results for very small size of dictionaries but better results for dictionaries bigger than 500 visual words, which are the commonly used configurations in BoVW approaches. Each layer of graph words performs much better than the SURF upon which they are built. The introduction of the topology in our features have a significant impact on the recognition performance using the same set of SURF features.

The average performance hides however some differences in the performance of each feature on some specific objects. To illustrate this we select two object categories where graph features and SURF give different performances in Figure **6** and Figure **7**. For the object "banana" from SIVAL, the isolated SURF features outperform the graph approach, see Figure **6**, whereas for the "Faces" category from Caltech-101 the graphs features perform better, see Figure **7**. This unequal discriminative power of each layer leads naturally to the use of the combination of the different layers in a single visual signature.

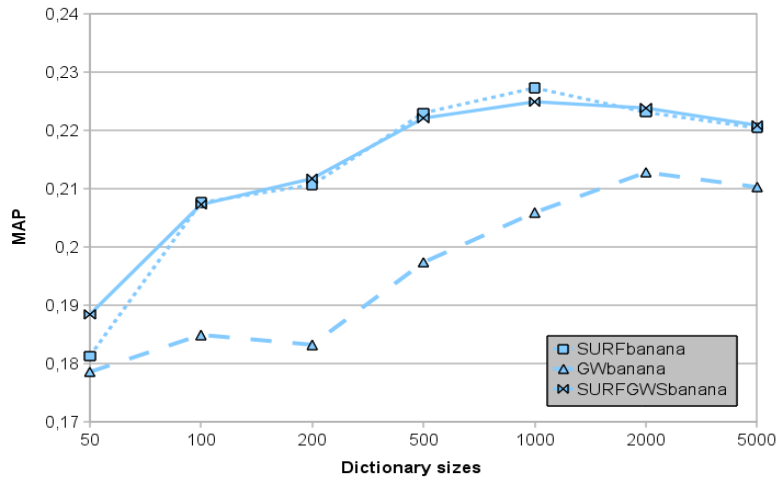

**Fig. 6.** MAP for the object "banana" from SIVAL where isolated SURF features (dotted curves) outperforms graphs (dashed curves). The multi-layer approach is the solid curve.

### 5.2   The multi-layer approach

The combination of graphs and SURF features upon which the graphs have been built is done by the concatenation of the signatures of each layer. The three curves in solid lines in Figure **5** correspond to the multi-layer approach using only the two bottom layers (SURF + 3 nearest neighbors graphs) in red, the three bottom layers (SURF + 3 nearest neighbors graphs + 6 nearest neighbors) in green and all the layers in blue. The improvement in the average MAP is clear, and each addition of layer improves the results. The average performance of the combination always outperforms the performance of each layer taken separately. For Caltech-101, the average MAP values of all methods are much lower which is not surprising as there are much more categories and images. Single layer of graphs gives results in the range 0.050-0.061 while the classical BoVW framework on SURF features performances are within 0.057-0.073 of average MAP values. The combination of all layers outperforms here again SURF or graphs used separately with average MAP values in the range of 0.061-0.077. The detailed results presented in Figure **6** and Figure **7** show that the combination of the visual signatures computed on each layer separately performs better or at least as well as the best isolated feature.

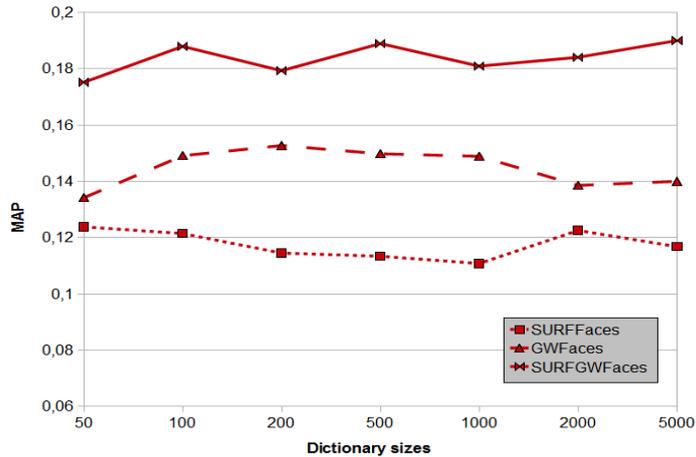

**Fig. 7.** MAP for category "Faces" from Caltech-101 where graphs (dashed curves) outperforms isolated SURF features (dotted curves). The multi-layer approach is the solid curves.

## 6   Conclusion and perspectives

In this paper, we have presented new graph features built upon SURF points as nodes and expressing spatial relations between local key points. The multi-layer approach using growing neighborhoods in several layers enables to capture the most discriminative visual information for different types of objects. Using growing spatial neighborhood clearly improves the results while each layer taken separately yields smaller improvements. Moreover, this approach introduces spatial information within the features and is therefore complementary and compatible with other recent improvements of the BoW framework for tacking geometry into account, such as the Spatial Pyramid Matching.

The future of this work is the application of the method to the recognition of objects in videos. The approach could be enhanced by refining some steps of the graphs construction and comparison. For instance, the selection of seeds could be performed by an adaptive method and the topology matrix be defined with a soft connectivity. In order to be efficient when processing a large amount of images, i.e. in videos, a graph embedding procedure could be applied to use an indexing structure that would speed up the recognition process.

## 7    Acknowledgments

This work is partly supported by a grant from the ANR (Agence Nationale de la Recherche) with reference ANR-09-BLAN-0165-02, within the IMMED project.

## 8    References

[1] D. G. Lowe, "Distinctive image features from scale-invariant keypoints," *International Journal of Computer Vision*, vol. 60(2), pp. 91-110, 2004.

[2] J. Sivic and A. Zisserman, "Video google: a text retrieval approach to object matching in videos," *ICCV'2003*, vol. 2, pp. 1470-1477, 2003.

[3] H. Bay, A. Ess, T. Tuytelaars, and L. V. Gool, "Surf:Speeded up robust features," *Computer Vision and Image Understanding*, vol. 110, pp. 346-359, 2008.

[4] K. Grauman and T. Darrell, "The pyramid match kernel: Discriminative classification with sets of image features," *In Proc. ICCV*, 2005.

[5] S. Lazebnik, C. Schmid, and J. Ponce, "Beyond bags of features: Spatial pyramid matching for recognizing natural scene categories,". *In Proc. CVPR*, 2006.

[6] A. Bosch, A. Zisserman, and X. Munoz, "Representing shape with a spatial pyramid kernel,". In *Proceedings of the 6th ACM international conference on Image and video retrieval* (CIVR '07). ACM, New York, NY, USA, 401-408.

[7] R. Albatal, P. Mulhem, Y. Chiaramella, "Visual Phrases for automatic images annotation," *CBMI'10*, Grenoble, France, 2010.

[8] A. Mahboubi, J. Benois-Pineau, D. Barba,"Joint tracking of polygonal and triangulated meshes of objects in moving sequences with time varying content," *IEEE International Conference on Image Processing*, vol. 2, pp. 403-406, 2001.

[9] H. Sahbi, J.-Y. Audibert, J. Rabarisoa, R. Keriven, "Robust matching and recognition using context-dependent kernels," *Proceedings of the 25th International Conference on Machine Learning*, pp. 856-863, 2008.

[10] P. H. Gosselin, M. Cord, S. Philipp-Foliguet, "Combining visual dictionary, kernel-based similarity and learning strategy for image category retrieval," *Computer Vision and Image Understanding*, Vol. 100(3), June 2008.

[11] SIVAL Data set: http://accio.cse.wustl.edu/sg-accio/SIVAL.html

[12] L. Fei-Fei, R. Fergus and P. Perona. "One-Shot learning of object categories". IEEE Trans. *Pattern Recognition and Machine Intelligence*.